\documentclass[aip,pop,reprint,superscriptaddress]{revtex4-1}

\usepackage[final]{graphicx}
\usepackage{amsmath}
\usepackage{verbatim}

\usepackage{graphicx} 
\usepackage{bm} 
\usepackage{dcolumn}

\usepackage{subfigure}
\usepackage{sidecap}

\usepackage[ 
colorlinks,
linkcolor=blue,
anchorcolor=blue,
citecolor=blue,
urlcolor=blue]
{hyperref}

\begin{document}

\preprint{}

\title{Excitation of nonlinear ion acoustic waves in CH plasmas}

\author{Q. S. Feng} 
\affiliation{HEDPS, Center for
	Applied Physics and Technology, Peking University, Beijing, 100871, China}

\author{C. Y. Zheng} \email{zheng\_chunyang@iapcm.ac.cn}
\affiliation{HEDPS, Center for
	Applied Physics and Technology, Peking University, Beijing, 100871, China}
\affiliation{Institute of Applied Physics and Computational
	Mathematics, Beijing, 100094, China}
\affiliation{Collaborative Innovation Center of IFSA (CICIFSA), Shanghai Jiao Tong University, Shanghai 200240, China}

\author{Z. J. Liu} 
\affiliation{HEDPS, Center for
	Applied Physics and Technology, Peking University, Beijing, 100871, China}
\affiliation{Institute of Applied Physics and Computational
	Mathematics, Beijing, 100094, China}

\author{C. Z. Xiao}
\affiliation{HEDPS, Center for
	Applied Physics and Technology, Peking University, Beijing, 100871, China}

\author{Q. Wang}
\affiliation{HEDPS, Center for
	Applied Physics and Technology, Peking University, Beijing, 100871, China}
	
\author{X. T. He}
\affiliation{HEDPS, Center for
	Applied Physics and Technology, Peking University, Beijing, 100871, China}
\affiliation{Institute of Applied Physics and Computational
	Mathematics, Beijing, 100094, China}
\affiliation{Collaborative Innovation Center of IFSA (CICIFSA), Shanghai Jiao Tong University, Shanghai 200240, China}

\date{\today}

\begin{abstract}

 Excitation of nonlinear ion acoustic wave (IAW) by an external electric field is demonstrated by Vlasov simulation. The frequency calculated by the  dispersion relation with no damping is verified much closer to the resonance frequency of the small-amplitude nonlinear IAW than that calculated by the linear dispersion relation. When the wave number $ k\lambda_{De} $ increases, the linear Landau damping of the fast mode (its phase velocity is greater than any ion's thermal velocity) increases obviously in the region of $ T_i/T_e < 0.2 $ in which the fast mode is weakly damped mode. As a result, the deviation between the frequency calculated by the linear dispersion relation and that  by the dispersion relation with no damping becomes larger with $k\lambda_{De}$ increasing. When $k\lambda_{De}$ is not large, such as $k\lambda_{De}=0.1, 0.3, 0.5$, the nonlinear IAW can be excited by the driver with the linear frequency of the modes. However, when $k\lambda_{De}$ is large, such as $k\lambda_{De}=0.7$, the linear frequency can not be applied to exciting the nonlinear IAW, while the frequency calculated by the dispersion relation with no damping can be applied to exciting the nonlinear IAW.

\end{abstract}


\maketitle

\section{\label{Sec: Introduction}Introduction}
Laser scattering involving stimulated Brillouin scattering (SBS), i.e., the coupling of a large amplitude light wave into a scattered light wave plus an ion acoustic wave (IAW), \cite{Kruer} play an important role in the successful ignition goal of inertial confinement fusion (ICF) \cite{(1), (9)_37}. 
 Many methods have been used to reduce the SBS \cite{(9)_49, (9)_46} scattering level, such as increasing the Landau damping of ion-acoustic waves\cite{Liu_8} and the saturation of ion-acoustic wave \cite{(9)_6,(9)_7}.  
Various mechanisms have been put forward to explain ion-acoustic wave saturation, including frequency detuning induced by particle-trapping,\cite{(9)_6_2,(9)_6_4,(9)_6_5} increased linear Landau damping due to kinetic ion heating,\cite{(9)_6_9,(9)_6_10} nonlinear damping induced by wave-breaking and trapping\cite{(9)_6_11,(9)_6_12} or coupling with higher harmonics\cite{(9)_6_13,(9)_6_14}. 

Multiple ion species are contained in the laser fusion program. The presence of multiple ion species can add additional branches to the IAW dispersion relation and change the total Landau damping significantly, which may provide the possibility of controlling SBS by increasing the linear damping of the IAW.\cite{Liu_8} When the external driving electric field is on, the trapping of particles will reduce Landau damping, if the time of the driver is long enough, the Landau damping will be nearly zero.\cite{(3)} At the same time, the dispersion of the small amplitude nonlinear IAW will be altered. The linear kinetic theory of ion acoustic waves in two ion species plasmas was researched by E. A. Williams et al. in detail.\cite{(7)} There, $k\lambda_{De}=0.1$ was taken, the linear frequency of the IAW was thought to be the resonance frequency of the IAW. In this paper, we can see when $k\lambda_{De}=0.1$, the linear frequency of the IAW was close to the frequency of the small amplitude nonlinear IAW (discussed later). In 2013, T. Chapman et al. researched the kinetic nonlinear frequency shift (KNFS) of the nonlinear ion acoustic wave in CH plasmas.\cite{T. Chapman_PRL}  The fundamental frequency $\omega_0$ for Vlasov results was given by simulation. They took the frequency of the IAW of zero amplitude by extrapolating from the lowest measured amplitude cases, i. e., $\omega_{0}=\omega(\phi\rightarrow0)$. In 2001, Rose and Russell\cite{Rose_2001POP} derived the expressions for the nonlinear dielectric function of Langmuir waves, thus the nonlinear dispersion relation. Then on this basis, in 2003, Rose presented the estimates to the spatial gain rate coefficients of the backward stimulated Raman scatter (BSRS) and backward stimulated Brillouin scatter (BSBS) due to trapping effects.\cite{Rose_2003POP} And also, Strozzi et al. \cite{Strozzi_2007POP} paid attention to the linear modes with distributions modified by electron trapping and researched the kinetic effects such as electron trapping in stimulated Raman backscatter by one-dimensional Eulerian Vlasov-Maxwell simulations. Then, they presented a framework for estimating when electron trapping nonlinearity was expected to be important in Langmuir waves.\cite{Strozzi_2012POP} In the work of Rose and Russell,\cite{Rose_2001POP} the finite wave amplitude $\phi$ was considered and the real part of the nonlinear dielectric function determined the nonlinear resonance condition. When the wave amplitude was small enough, the solutions in the limit $\phi\rightarrow0$ corresponded to the infinitesimal amplitude Bernstein-Greene-Kruskal (BGK) modes\cite{BGK}, which was consistent to that found by Holloway and Dorning.\cite{Holloway_1991} Their results were for the electron plasma waves while in this paper the resonance condition is applied to IAWs in multi species plasmas especially for the infinitesimal amplitude BGK modes. In this paper, the dispersion relation with no damping, corresponding to the real part of the linear dielectric function with no damping, is given to calculate the fundamental frequency $\omega_0$ of the small-amplitude nonlinear IAW which is undamped plasma wave\cite{Holloway_1991}. The electrons and ions are all kinetic, but the Landau damping of the nonlinear IAW can be ignored as a result of particle trapping. We can see when $k\lambda_{De}$ is not large, the deviation between the frequency calculated by the linear dispersion relation (defined as the linear frequency) and that calculated by the dispersion relation with no damping of the infinitesimal amplitude nonlinear IAW is close to each other, but with $k\lambda_{De}$ increasing, the deviation becomes larger. By Vlasov simulation, the frequency calculated by the dispersion relation with no damping is verified much closer to the resonance frequency of the small-amplitude nonlinear IAW.

This paper discusses the change of the linear Landau damping rate of the fast IAW mode (the phase velocity of which is faster than any ion's thermal velocity) and the slow IAW mode (the phase velocity of which is close to the thermal velocity of one of the plasma species) in CH plasmas when $ k\lambda_{De} $ varies. When $k\lambda_{De}$ increases, the linear Landau damping of the fast mode will increase obviously in the region of $T_i/T_e\lesssim0.2$. However, this result is for the linear situation. When the driving electric field is on, the distribution of the particles will keep flat at the phase velocity of the IAW, thus turning off the linear Landau damping. As a result, the deviation between the frequency calculated by the linear dispersion relation and that by the dispersion relation with no damping becomes larger with $k\lambda_{De}$ increasing.
When $k\lambda_{De}$ is not large, the nonlinear IAW can be excited by the driver with the linear frequency of the modes, which indicates that the frequency of the nonlinear IAW can be approximated by the linear dispersion relation such as $k\lambda_{De}=0.1, 0.3, 0.5$. However, when $k\lambda_{De}=0.7$, the linear dispersion relation can not be applied to exciting the nonlinear IAW, while the dispersion relation with no damping can be applied to exciting the nonlinear IAW.

This paper is organized as follows:
In Sec. \ref{Sec:Theory analysis}, the linear dispersion relation and the dispersion relation with no damping are presented. The change of the phase velocity and Landau damping of the IAW modes with $k\lambda_{De}$ is discussed. And the dispersion relation with no damping is provided to calculate the frequency of the small-amplitude nonlinear IAW in CH plasmas. 
In Sec. \ref{Sec:Vlasov simulation},  the excitation of the nonlinear ion-acoustic wave by an external electric field with the linear frequency and the frequency calculated by the dispersion relation with no damping is demonstrated by one-dimensional (1D) Vlasov simulation. And the Vlasov simulation results of the excitation of the nonlinear IAW both for the fast mode and the slow mode are shown when $k\lambda_{De}$ varies.
In Sec. \ref{Sec:Discussion}, the discussion of the effect of the nonlinear frequency shift is given.
In Sec. \ref{Sec:Summary}, our main results are summarized. 

\section{\label{Sec:Theory analysis}Theoretical analysis}
The linear kinetic theory of ion acoustic wave in a non-magnetized, homogeneous plasma consisting of multi-species ions is considered. Considering a neutral, fully ionized, unmagnetized CH plasmas (1:1 mixed) with the same temperature of all ion species ($T_H=T_C=T_i$), the linear frequencies of the plasmas are then given by the zeros of the plasma dielectric function, i.e., $\epsilon(k,\omega)=0$, which gives the linear dispersion relation of the ion acoustic wave in multi-ion species plasmas
\begin{equation}
\label{Eq:Dispersion}
	\epsilon_L(\omega,k)=1+\chi_e+\sum_\beta \chi_{i\beta}=0,
\end{equation}
where $\chi_e$ is the susceptibility of electron and $\chi_{i\beta}$ is the susceptibility of species $\beta$ (H ion or C ion). $\chi_j$ (j present for electron, H ion or C ion) can be expressed by $Z$ function
\begin{equation}
\label{Eq:Chi}
\chi_j=\frac{1}{(k\lambda_{Dj})^2}(1+\xi_jZ(\xi_j)),
\end{equation}
where a Maxwellian velocity distribution for all  species is assumed. $\xi_j=\omega/(\sqrt{2}kv_{tj})$ is generally complex, $\omega=\omega_r+i\gamma$, $k$ is the frequency and the wave number of the given mode (such as the fast or slow IAW mode). $v_{tj}=\sqrt{T_j/m_j}$ ($T_j$, $m_j$ are the temperature and the mass of particle $j$) is the thermal velocity of particle $j$. $\lambda_{Dj}$ is the Debye length $\lambda_{Dj}=\sqrt{T_j/4\pi n_jZ_j^2e^2}$, i.e., $\lambda_{Dj}=v_{tj}/\omega_{pj}$ ($\omega_{pj}=\sqrt{4\pi n_jZ_j^2e^2/m_j}$ is the plasma frequency for specie j). $Z$ function is the dispersion function 
\begin{equation}
\label{Eq:Zeta}
\begin{aligned}
Z(\xi_j)  &     =\frac{1}{\sqrt{\pi}}\int_{-\infty}^{+\infty}\frac{e^{-v^2}}{v-\xi_j}dv\\
& =i\sqrt{\pi}e^{-\xi_j^2}(1+erf(i\xi_j)),
\end{aligned}
\end{equation}
in which $erf(i\xi_j)=2/\sqrt{\pi}\int_0^{i\xi_j} e^{-\eta^2}d\eta$ is the error function, $\xi_j$ is a complex variable. The direct numerical solution to the Eq. (\ref{Eq:Dispersion}) can be solved by Newton-Raphson iterative method. The newly developed accurate algorithm of the Faddeyeva (plasma dispersion) function $\omega(z)=e^{-z^2}(1+erf(iz))$ (where $z=x+iy$ is a complex variable) is showed by M. R. Zaghloul\cite{ACM}.  

\begin{figure}[!tp]
	\includegraphics[width=1.0\columnwidth]{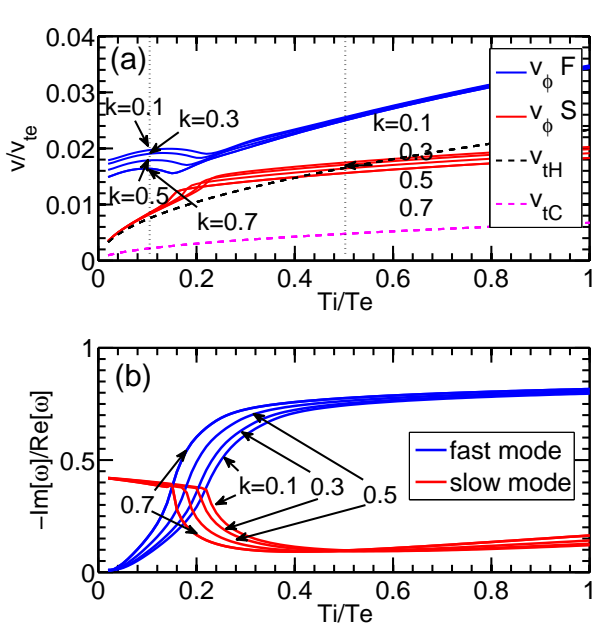}
	
	\caption{\label{Fig:LandauDamping}(Color online)  (a) The phase velocity and (b) the linear Landau damping of the fast mode and the slow mode as a function of $T_i/T_e$ when $k\lambda_{De}=0.1, 0.3, 0.5, 0.7$. }
\end{figure}

\begin{table*}
	\caption{\label{table1} The main results of the  linear dispersion relation and dispersion relation with no damping}
	\begin{ruledtabular}
		
		\begin{tabular}{|c|cccc|cccc|}

			\hline
			& \multicolumn{4}{c|}{\bf{fast mode}}&\multicolumn{4}{c|}{\bf{slow mode}}\\
			\hline
			$\bf{T_i/T_e}$& \multicolumn{4}{c|} {{\bf0.1}} & \multicolumn{4}{c|}{{\bf0.5}}  \\
			
			\hline
			$\bf{k\lambda_{De}}$& {\bf0.1} & {\bf0.3} &{\bf 0.5} & {\bf0.7}
			& {\bf0.1} & {\bf0.3} & {\bf0.5} & {\bf0.7} \\
			\hline

			$\bf{Re(\omega_L)/10^{-3}\omega_{pe}}$& ${\bf 1.965}$&${\bf5.700}$& ${\bf8.930}$   &${\bf 11.47}$& $\bf1.746$& $\bf5.121$
			&$\bf8.250$  &$\bf10.91$  \\
			
			$\bf{\omega_N/10^{-3}\omega_{pe}}$& ${\bf 1.912}$& ${\bf5.505}$& ${\bf 8.478}$ &${\bf 10.60}$&$\bf1.746$ &$\bf5.094$
			& $\bf8.079$   &$\bf10.60$  \\
			
			$\bf{|Im(\omega_L)|/10^{-3}\omega_{pe}}$& {0.1561}  & {0.5150}& {1.018}& {1.767}& 0.1707 & 0.4857
			& 0.7682& 1.055 \\
			
			$\bf{|Im(\omega_L)/Re(\omega_L)|}$& {0.07946}  & {0.09035}& {0.114}& {0.154}& 0.09775  & 0.09485
			& 0.09312& 0.09667 \\

			\hline
			
		\end{tabular}
		
	\end{ruledtabular}
\end{table*}

Figs. \ref{Fig:LandauDamping}(a) and \ref{Fig:LandauDamping}(b) show the phase velocity and the linear Landau damping rate of the fast mode and the slow mode in the condition of $k\lambda_{De}=0.1, 0.3, 0.5, 0.7$. With $k\lambda_{De}$ increasing, the phase velocity of the fast mode will decrease and be closer to the thermal velocity of H ions especially in the region of $T_i/T_e\lesssim0.2$ in which the Landau damping rate of the fast mode is weak. Thus, the Landau damping rate of the fast mode increases obviously with $k\lambda_{De}$ increasing especially in the region of $T_i/T_e\lesssim0.2$.\\

When the external driving electric field is turned on, the total electric potential of the system including the electric field of the driver and the electric field of IAW will trap particles, thus making the distribution flat at the wave phase velocity, which will reduce Landau damping.\cite{(3)} If the duration time of the driver is long enough, the Landau damping will be turned off, i.e.,  $Im(\omega)\approx0$. By retaining only the real part of Eq. (1), the dispersion relation with no damping of the infinitesimal amplitude nonlinear modes (described briefly below as \textquotedblleft the dispersion relation with no damping\textquotedblright ) is given by
\begin{equation}
\label{Eq:Nonlinear}
Re(\epsilon_L(Re(\omega), k))=0,
\end{equation}
in which $Re(\omega)$ is the frequency of the infinitesimal amplitude nonlinear mode by taking $Im(\omega)=0$. $\epsilon_L$ is defined in Eq. (1). Where Eq. (\ref{Eq:Nonlinear}) describes only small-amplitude nonlinear IAW. A Maxwellian distribution for all species is used and the width of the plateau, where $\partial f_0/\partial v=0$, is infinitesimal. 

\begin{figure}[!tp]
	\includegraphics[width=\columnwidth]{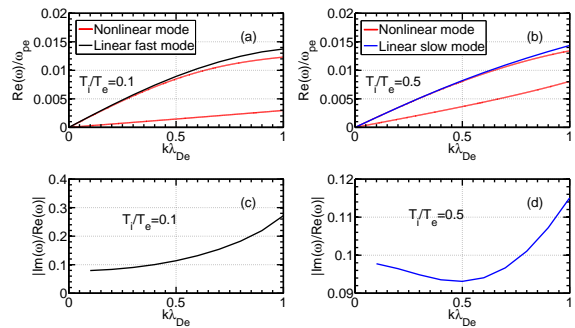}

	\caption{\label{Fig:LinearNonlinear}(Color online) The dispersion relations of the fast mode and the slow mode calculated by linear dispersion relation and dispersion relation with no damping in the condition of (a) $T_i/T_e=0.1$, (b) $T_i/T_e=0.5$. Note that \textquotedblleft Nonlinear mode" presents for the \textquotedblleft infinitesimal amplitude nonlinear mode" (red lines in (a) and (b)). The linear Landau damping of (c) the fast mode in the condition of $T_i/T_e=0.1$ and (d) the slow mode in the condition of $T_i/T_e=0.5$.}
\end{figure}

 As $k\lambda_{De}$ increases, the linear Landau damping of the fast mode increases obviously as shown in Fig. \ref{Fig:LinearNonlinear}(c), therefore, the frequency of the linear fast mode deviates from that of the nonlinear IAW mode (Fig. \ref{Fig:LinearNonlinear}(a), the upper branch of the red line). When $k\lambda_{De}=0.7$, the deviation between the frequency of the linear fast mode and the nonlinear mode, which is calculated by the dispersion relation with no damping, is nearly as large as $7\%$. The same analysis of the slow mode in the condition of $T_i/T_e=0.5$ is given in Figs. \ref{Fig:LinearNonlinear}(b) and \ref{Fig:LinearNonlinear}(d). As shown in Fig. \ref{Fig:LinearNonlinear}(d), the linear Landau damping rate $|Im(\omega)/Re(\omega)|$ of the slow mode decreases first and then increases as $k\lambda_{De}$ increases. The deviation between the frequency of the linear slow mode and the nonlinear IAW mode (Fig. \ref{Fig:LinearNonlinear}(b), the upper branch of the red line) becomes larger as $k\lambda_{De}$ increases as shown in Fig. \ref{Fig:LinearNonlinear}(b). When $k\lambda_{De}=0.7$, the deviation is nearly $3\%$. For $k\lambda_{De}=0.1, 0.3, 0.5$, the frequency of the linear fast mode in the condition of $T_i/T_e=0.1$ and the linear slow mode in the condition of $T_i/T_e=0.5$ are near the frequency of the nonlinear IAW mode as shown in Figs. \ref{Fig:LinearNonlinear}(a) and \ref{Fig:LinearNonlinear}(b), so the nonlinear IAW can be excited by the driver with the linear frequency of the IAW.

These cases are calculated by linear dispersion relation and dispersion relation with no damping to get the frequency of the IAW ${Re(\omega_L)}$, $\omega_N$, the linear Landau damping $|Im(\omega_L)|$, the relative linear Landau damping rate $|Im(\omega_L)/Re(\omega_L)|$ of the fast mode in the condition of $T_i/T_e=0.1$ and the slow mode in the condition of $T_i/T_e=0.5$. The results of the linear dispersion relation and the dispersion relation with no damping of different $k\lambda_{De}$ are shown in Table I. As shown in Fig. \ref{Fig:LandauDamping}(b), when $T_i/T_e=0.1$, the less damped mode is the fast mode, when $T_i/T_e=0.5$, the less damped mode is the slow mode. In this paper, only the weakly damped modes are considered, i.e., a fast mode in the condition of $T_i/T_e=0.1$ and a slow mode in the condition of $T_i/T_e=0.5$ are taken as representations.

\section{\label{Sec:Vlasov simulation}Vlasov simulation}

In the one-dimensional (1D) Vlasov code,\cite{Liu-2009POP,Liu-2009POP_1} assuming that the external driving electric field $E_d$ is along the x direction. For particle specie s (s presents for electrons, H ions or C ions in this paper), then the motion of specie s can be described by Vlasov-Poisson equations:
\begin{equation}
\label{Eq:Vlasov}
\frac{\partial f_s}{\partial t}+v_{xs}\frac{\partial f_s}{\partial x}+\frac{q_s}{m_s}(E_x+E_d)\frac{\partial f_s}{\partial v_x}=0,
\end{equation}

\begin{equation}
\label{Eq:Poisson}
\frac{\partial E_x}{\partial x}=4\pi \sum\limits_{s}n_{s0}q_s\int\limits_{-\infty}^{+\infty}f_sdv,
\end{equation}
where all of the particles including electrons and ions are taken as kinetic particles and $f_s$ is the normalized distribution function for specie s. The initial equilibrium distribution function for each specie particles $f_{s0}$ is normalized Maxwellian distribution satisfying $\int_{-\infty}^{+\infty}f_{s0}dv=1$.  $E_x$ is the self-consistent electrostatic field in plasmas and is calculated from Poisson's equation. And $q_s$, $m_s$ and $n_{s0}$ present for the charge, mass and the background equilibrium density of particle s.

To solve Vlasov equation (\ref{Eq:Vlasov}), we split the time-stepping operator into free-streaming in x and motion in $v_x$,\cite{7_5-1976JCP,7-2004CPC} then we can get the advection equations:
\begin{equation}
\begin{aligned}
\label{Eq:split1}
&\frac{\partial f_s}{\partial t}+v_{xs}\frac{\partial f_s}{\partial x}=0,\\
\end{aligned}
\end{equation}

\begin{equation}
\begin{aligned}
\label{Eq:split2}
&\frac{\partial f_s}{\partial t}+\frac{q_s}{m_s}(E_x+E_d)\frac{\partial f_s}{\partial v_x}=0,
\end{aligned}
\end{equation}
a third order Van Leer scheme (VL3)\cite{VL3, 10-2006POP} is taken to solve the advection equations (\ref{Eq:split1}) and (\ref{Eq:split2}). To solve the Poisson equation (\ref{Eq:Poisson}), fast Fourier transform (FFT) method\cite{1965FFT, FFTsolvePoisson} is taken to calculate the electrostatic field.

The excitation of the nonlinear IAW is simulated by 1D1V Vlasov code. In this code, the electrons and ions are all kinetic. To solve the electrons and ions kinetic behavior, the phase space domain is $[0, L_x]\times[-v_{max}, v_{max}]$, where $L_x=2\pi/k$ is the longest wavelength that fits into the simulation box and $v_{max}=8v_{tj}$, i.e., for kinetic electrons $v_{max}=8v_{te}$, for kinetic ions $v_{max}=8v_{ti}$ ($i$ presents for H ions or C ions in this paper). The phase space is discretized with $N_x=128$ grid points in the spatial domain and $N_v=256$ in the velocity domain. The periodic boundary condition is taken in the spatial domain. The time step is $dt=0.1\omega_{pe}^{-1}$. The envelope of the external driver is given by
\begin{equation}
\tilde{E}_d(t)=\tilde{E}_d\frac{1}{1+(\frac{t-t_0}{\frac{1}{2}t_0})^{10}},
\end{equation}
where $\tilde{E}_d=eE_d\lambda_{De}/T_e$ is the maximum amplitude of the external driver, tilde presents for normalization. $t_0$ is the duration time of the peak electric field.
To generate a driven IAW, one thus considers
\begin{equation}
\tilde{E}_d(x,t)=\tilde{E}_d(t)sin(kx-\omega_d t),
\end{equation}
where $\omega_d, k$ are the frequency and the wave number of the external driver. In this paper, to excite the nonlinear IAW, $\omega_d$ is chosen as the  frequency of the modes calculated by linear dispersion relation or dispersion relation with no damping.

\subsection{\label{Subsec:A. Response of IAW}Response of IAW to the driver}

\begin{figure}[htd]
\includegraphics[width=\columnwidth]{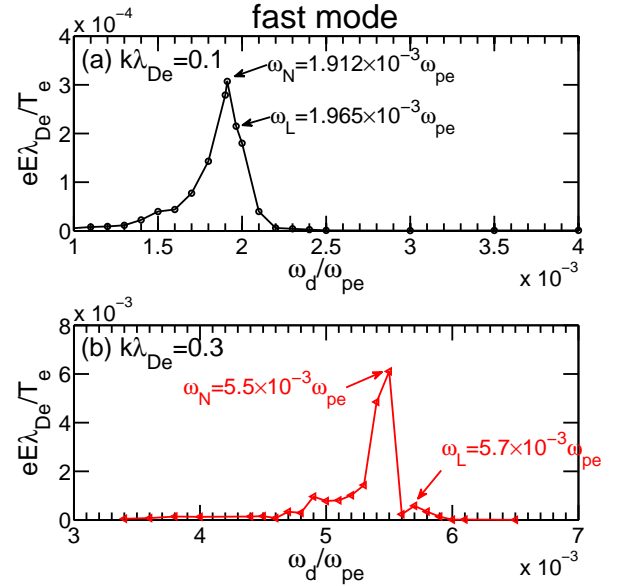}
	\caption{\label{Fig:Resonance}(Color online) The resonance curve of the fast mode ($T_i/T_e=0.1$) in the condition of (a) $k\lambda_{De}=0.1$, (b) $k\lambda_{De}=0.3$. The amplitude of the driver $eE_d\lambda_{De}/T_e=1\times10^{-3}$.}
\end{figure}

\begin{figure}[htd]
	\includegraphics[width=\columnwidth]{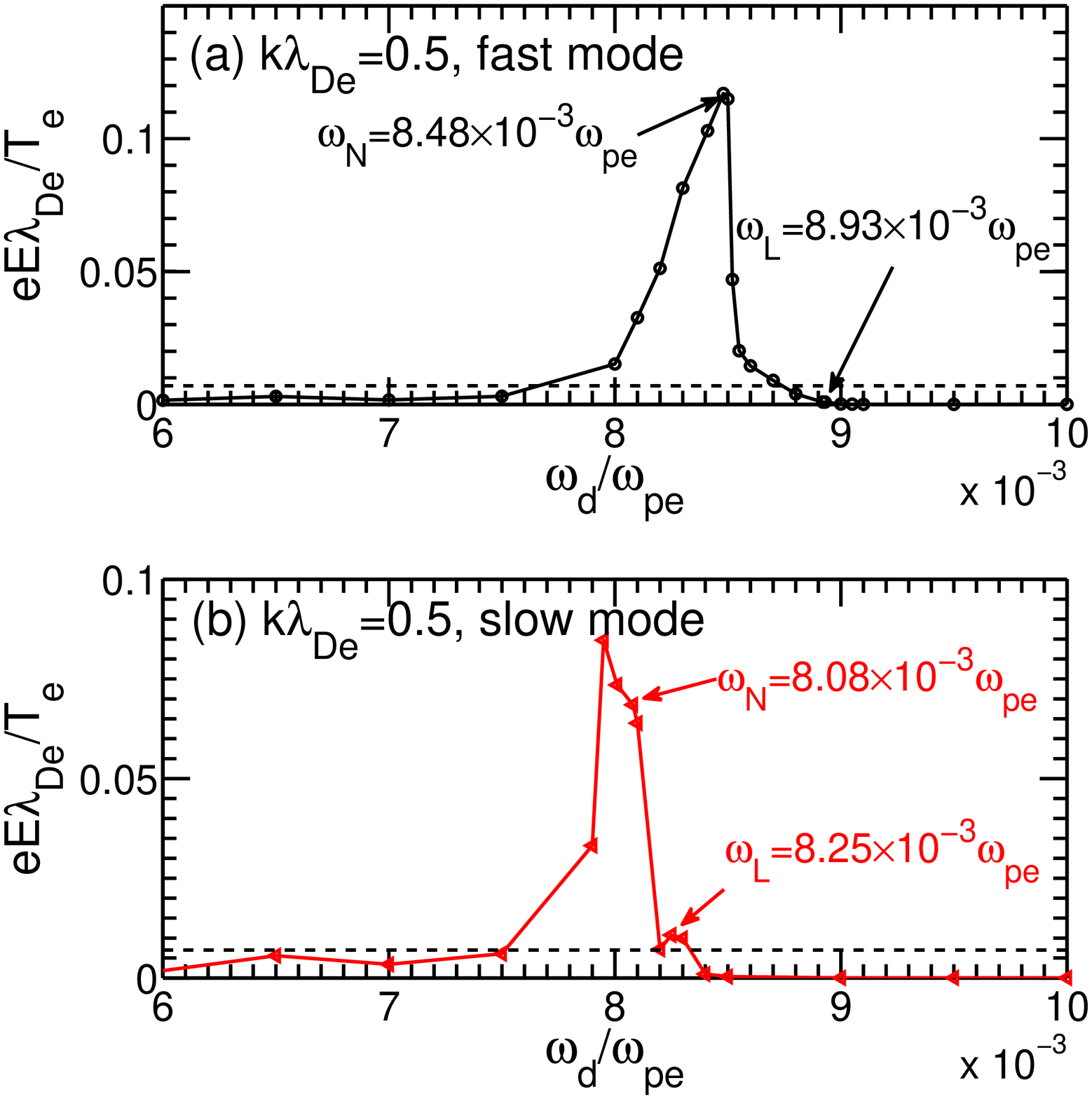}
	\caption{\label{Fig:Resonance2}(Color online) The resonance curve for (a) the fast mode, $T_i/T_e=0.1$, (b) the slow mode, $T_i/T_e=0.5$ in the condition of $k\lambda_{De}=0.5$, the black dashed line presents for the amplitude of the driver $eE_d\lambda_{De}/T_e=7\times10^{-3}$.}
\end{figure}

\begin{figure}[htd]
	\includegraphics[width=\columnwidth]{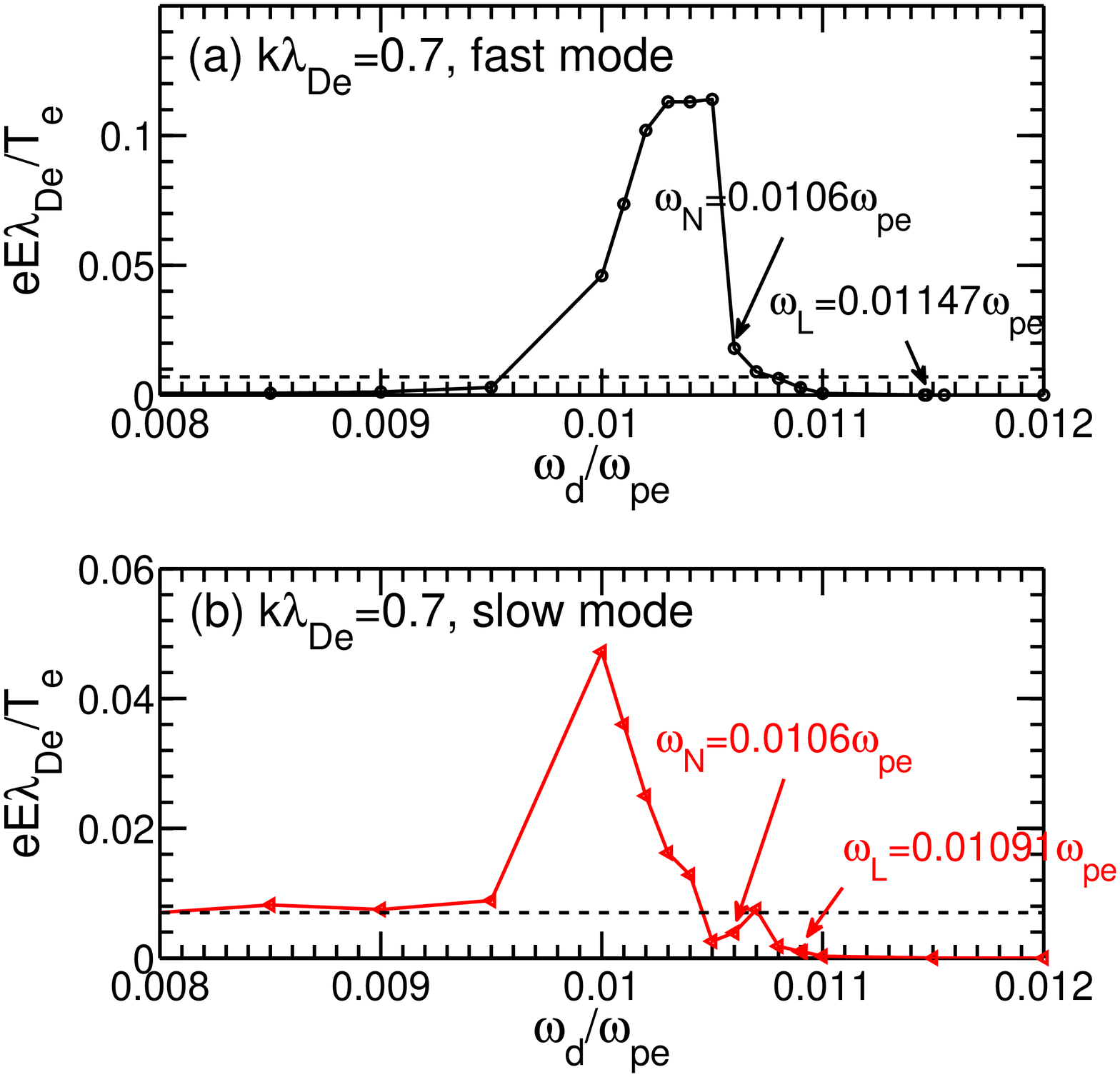}
	\caption{\label{Fig:Resonance3}(Color online) The resonance curve for (a) the fast mode, $T_i/T_e=0.1$, (b) the slow mode, $T_i/T_e=0.5$ in the condition of $k\lambda_{De}=0.7$, the black dashed line presents for the amplitude of the driver $eE_d\lambda_{De}/T_e=7\times10^{-3}$.}
\end{figure}

\begin{figure*}[!tp]
	\includegraphics[width=2\columnwidth]{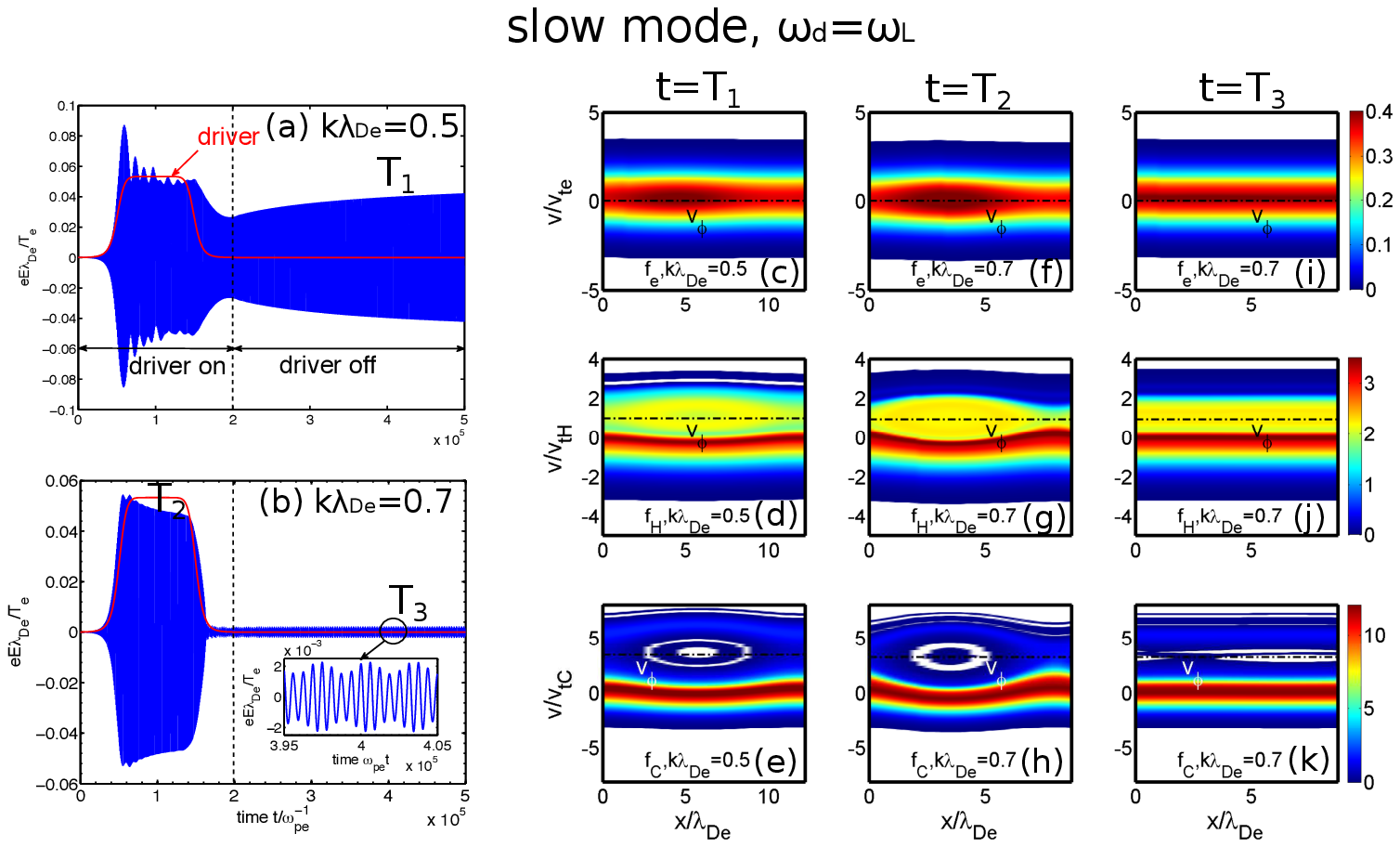}
	\caption{\label{Fig:SlowMode}(Color online) The total electric field and the envelope curve of the external driving electric field (driver) for the slow mode, $T_i/T_e=0.5$, in the condition of (a) $k\lambda_{De}=0.5$, (b) $k\lambda_{De}=0.7$. The phase picture of (c)-(e) electrons, (f)-(h) H ions, (i)-(k) C ions at the time of $T_1$ ($\sim4\times10^5\omega_{pe}^{-1}, k\lambda_{De}=0.5$), $T_2$ ($\sim1\times10^5\omega_{pe}^{-1}, k\lambda_{De}=0.7$), $T_3$ ($\sim4\times10^5\omega_{pe}^{-1}, k\lambda_{De}=0.7$). The driver is with the linear frequency of the slow mode calculated by the linear dispersion relation.}
	
\end{figure*}
To get the resonance curve, $k\lambda_{De}$ and other conditions are fixed, the frequency of the driver varies. The resonance curves of the fast mode ($T_i/T_e=0.1$)  in the condition of $k\lambda_{De}=0.1, 0.3$ are shown in Fig. \ref{Fig:Resonance}. The external driving electric field is turned on from the time of 0 to $2\times10^5\omega_{pe}^{-1}$ with the duration time $t_0=1\times10^5\omega_{pe}^{-1}$ of the maximum amplitude $eE_d\lambda_{De}/T_e=1.0\times10^{-3}$. When $k\lambda_{De}=0.1$, for the harmonics have much larger growth rate than the fundamental mode, the harmonics will carry a large part of energy of the IAW and prevent the fundamental mode from growing up. Then all modes will start to saturate and the IAW amplitude will saturate at a smaller level than the driver amplitude, which is similar to the results of the electron acoustic wave (EAW) researched by Xiao et al.\cite{Xiao_2014POP}. After $t=2\times10^5\omega_{pe}^{-1}$, the driver is turned off, the amplitude of the electric field of the driving IAW, $eE\lambda_{De}/T_e$ (or $eE_{IAW}\lambda_{De}/T_e$), is obtained at the time of $t\sim4\times10^{5}\omega_{pe}^{-1}$ (long after the driver has been turned off). As shown in Fig. \ref{Fig:Resonance}(a), in the condition of $k\lambda_{De}=0.1$ for the fast mode, $T_i/T_e=0.1$, the resonance frequency is near $\omega_N=1.912\times10^{-3}\omega_{pe}$ which is the frequency calculated by the dispersion relation with no damping rather than the frequency calculated by the linear dispersion relation $\omega_L=1.965\times10^{-3}\omega_{pe}$. The same results are obtained in the condition of $k\lambda_{De}=0.3$ for the fast mode ($T_i/T_e=0.1$) as shown in Fig. \ref{Fig:Resonance}(b). The resonance frequency is very close to the frequency calculated by the dispersion relation with no damping $\omega_N=5.505\times10^{-3}\omega_{pe}$ and the resonance peak is much larger than the amplitude of the driving IAW by the driver with the frequency calculated by the linear dispersion relation $\omega_L=5.7\times10^{-3}\omega_{pe}$.
When the nonlinear IAW excited by the driver is weak, the nonlinear frequency shift (NFS)\cite{(9)} of the IAW is small, thus the resonance peak is very close to the frequency calculated by the dispersion relation with no damping.

In the same way, when $k\lambda_{De}=0.5$, the resonance curves of the fast mode and the slow mode are given as shown in Fig. \ref{Fig:Resonance2}. It's also found that the resonance peak is close to the frequency calculated by the dispersion relation with no damping $\omega_N$. The linear frequency is higher but not much too higher than the resonance frequency, thus the IAW can be excited by the driver with the linear frequency especially when the driver amplitude is large. 

When $k\lambda_{De}=0.7$, as shown in Fig. \ref{Fig:Resonance3}, the resonance frequency is lower than the frequency calculated by the dispersion relation with no damping $\omega_N$ and also much lower than the linear frequency $\omega_L$. When $k\lambda_{De}$ is as large as 0.7, the harmonics are very weak and the fluid NFS from harmonic generation is negligible. It's caused by the negative nonlinear frequency shift due to the particles trapping when the IAW amplitude is large.\cite{T. Chapman_PRL, (9)} This phenomenon is common in the nonlinear IAW in the stimulated Brillouin scattering (SBS)\cite{Liu-2011POP} and other BGK modes such as nonlinear electron acoustic waves (EAWs)\cite{Valentini-2006POP}, Langmuir waves\cite{Rose_2001POP} and kinetic electrostatic electron nonlinear (KEEN) waves\cite{Johnston-2009POP}. For the linear frequency $\omega_L$ is far away from the resonance frequency, no matter how large of the driver amplitude, the IAW can nearly not be excited by the driver with the linear frequency.

\subsection{\label{Subsec:B. Large amplitude driver}Large amplitude driver with the linear frequency of the IAW}

\begin{figure*}[!tp]
	\includegraphics[width=2\columnwidth]{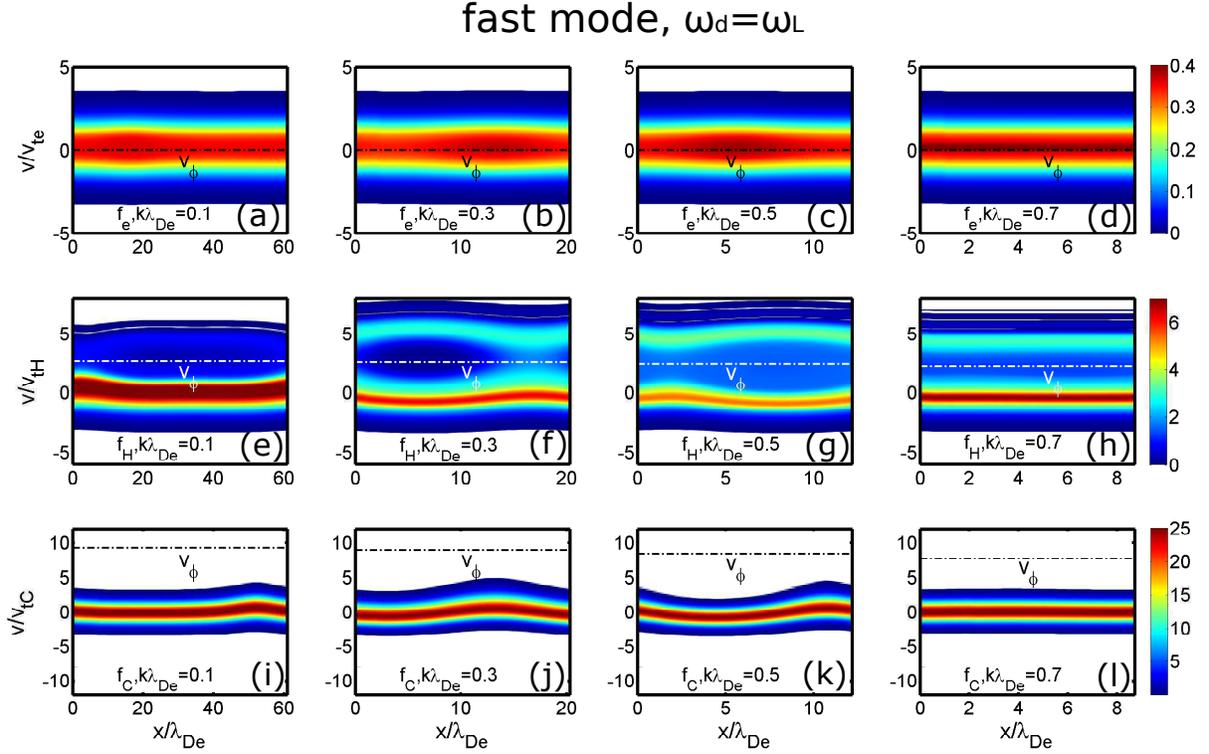}
	\caption{\label{Fig:FastMode}(Color online) The phase picture of (a)-(d) electrons, (e)-(h) H ions, (i)-(l) C ions for the fast mode ($T_i/T_e=0.1$) in the condition of $k\lambda_{De}=0.1, 0.3, 0.5, 0.7$ when the external driving electric field is off. The driver is with the frequency of the fast mode calculated by the linear dispersion relation.}
\end{figure*}
To show more clearly the difference of the excitation of the IAW and the phase pictures of particles in different $k\lambda_{De}$, the driver amplitude takes a large value $eE_d\lambda_{De}/T_e=5.33\times10^{-2}$.
As shown in Figs. \ref{Fig:SlowMode}(a) and \ref{Fig:SlowMode}(b), the external driving electric field (driver) is turned on from the time of 0 to $2\times10^5\omega_{pe}^{-1}$ with the duration time $t_0=1\times10^5\omega_{pe}^{-1}$ as the same. The frequency of the driver takes the linear frequency of the IAW as shown in Table I. For example, in Fig. \ref{Fig:SlowMode}(a), $k\lambda_{De}=0.5$, $T_i/T_e=0.5$, so the linear frequency of the slow mode is $Re(\omega_s)=8.250\times10^{-3}\omega_{pe}$, in the same way, in Fig. \ref{Fig:SlowMode}(b), $k\lambda_{De}=0.7$, $T_i/T_e=0.5$, so the frequency of the driver is $1.091\times10^{-2}\omega_{pe}$. After the driver is turned off, the amplitude of the electric field in the system in the condition of $k\lambda_{De}=0.5$ is obviously larger than that in the condition of $k\lambda_{De}=0.7$.  
As shown in Fig. \ref{Fig:SlowMode}(a), in the condition of $k\lambda_{De}=0.5$, when the driver is turned on, the electric field amplitude may exhibit slow oscillations on the ion bounce time scale ($\tau_{bi}=2\pi/\sqrt{kq_iE/m_i}$) as the wave and resonant particles exchange energy. This process allows phase locking of the mode to the driver, which will make the distribution flat at the phase velocity thereby reducing Landau damping.\cite{(3)} After the driver is turned off, for the frequency of the driver is near the resonance frequency of the nonlinear IAW as shown in Fig. \ref{Fig:Resonance2}(b), the driver with the linear frequency will couple significant energy into  a BGK mode especially when the driver amplitude is large, thus the undriven BGK-like mode with relatively large amplitude will be established, known as auto-resonance which doesn't require feedback to maintain resonance. Thus, the trapping of electrons, H ions and C ions is very obvious as shown in Figs. \ref{Fig:SlowMode}(c)-\ref{Fig:SlowMode}(e) after the driver is turned off.
However, the driver is added into the CH plasmas systems in Fig. \ref{Fig:SlowMode}(b) as same as in Fig. \ref{Fig:SlowMode}(a). For Fig. \ref{Fig:SlowMode}(b), $k\lambda_{De}=0.7$, for the linear frequency $\omega_L$ is far away from the resonance frequency of the slow mode as shown in Fig. \ref{Fig:Resonance3}(b), the driver with the linear frequency $\omega_L$ can nearly not excite a large amplitude IAW by the way of resonance excitation no matter how large of the driver amplitude. As a result, after the driver is off, the IAW amplitude will fall down to a very small level and stabilize at a small but finite amplitude. In the same way, the small amplitude IAW will trap particles although the trapping width is very small, thus making the distribution of particles flat at the phase velocity thereby reducing Landau damping to nearly zero after several bounce periods. Therefore, the IAW stabilize at a very small but nonzero amplitude. This process allow the driver with the linear frequency $\omega_L$ to couple small but nonzero energy into a BGK mode. At the time of $T_3$ (as shown in Fig. \ref{Fig:SlowMode}(b), $T_3$ is the time after the diver is off in the case of $k\lambda_{De}=0.7$, $T_3\sim4\times10^{5}\omega_{pe}^{-1}$), the ion bounce periodic time $\tau_{bi}$ shows several times of the IAW period $\tau_{pi}=2\pi/\omega_{pi}$ when the internal IAW amplitude  $eE^{int}\lambda_{De}/T_e$ is as low as $2\times10^{-3}$. Figs. \ref{Fig:SlowMode}(f)-\ref{Fig:SlowMode}(h) show the phase picture of electrons, H ions and C ions at the time of $T_2$ when the driver is on. The trapping of the particles is obvious due to the interaction of the external driving electric field and particles when the driver is on. This process makes the distribution of particles flat at the phase velocity of IAW, $v_\phi$, thereby reducing the Landau damping. When the driver is off, the distribution keeps flat at $v_\phi$, but the trapping width of the particles at the time of $T_3$ (see Figs. \ref{Fig:SlowMode}(i)-\ref{Fig:SlowMode}(k)) is narrower than that at the time of $T_1$ (see Figs. \ref{Fig:SlowMode}(c)-\ref{Fig:SlowMode}(e), where $T_1$ is the time after the driver is off in the case of $k\lambda_{De}=0.5$, $T_1\sim4\times10^{5}\omega_{pe}^{-1}$) as a result of the lower amplitude of IAW electric field when the driver is off.

\begin{figure}[!tp]
	\includegraphics[width=\columnwidth]{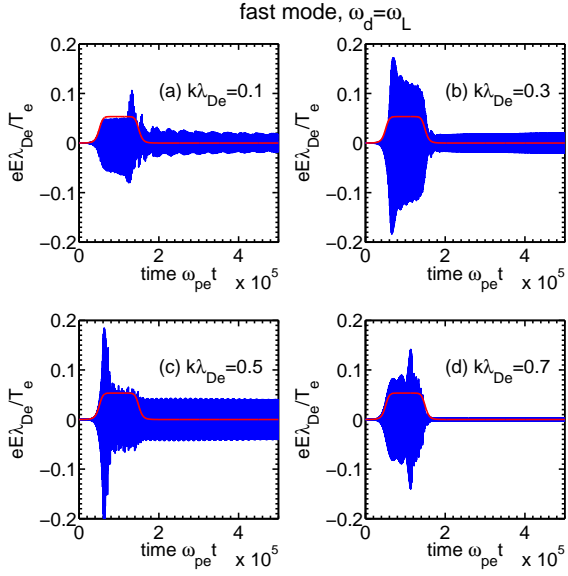}
	\caption{\label{Fig:Electric_FastMode}(Color online) The total electric field of the system (blue line) and the envelope of the external driving electric field (driver, red line) for the fast mode ($T_i/T_e=0.1$) in the condition of (a) $k\lambda_{De}=0.1$, (b) $k\lambda_{De}=0.3$, (c) $k\lambda_{De}=0.5$, (d) $k\lambda_{De}=0.7$, which is corresponding to Fig. \ref{Fig:FastMode}. The driver is with the linear frequency of the fast mode calculated by the linear dispersion relation. }
\end{figure}

Fig. \ref{Fig:FastMode} shows the trapping of the particles including electrons, H ions and C ions in the cases of $k\lambda_{De}=0.1$, 0.3, 0.5, 0.7 for the fast mode when the driver is off. The electric fields of the system are shown in Fig. \ref{Fig:Electric_FastMode} (discussed later), where the maximum amplitude of the driver $eE_d\lambda_{De}/T_e=5.33\times10^{-2}$ is taken as an example and the driver frequency takes the linear frequency of the IAW. Here, $v_\phi$, the phase velocity of the modes, are marked in the phase pictures as shown in Fig. \ref{Fig:FastMode}. For the phase velocity of the fast mode, $v_{\phi f}$, is much larger than the thermal velocity of C ions, $v_{tC}$, C ions can not be trapped in all cases for the fast mode (see Figs. \ref{Fig:FastMode}(i)-\ref{Fig:FastMode}(l)), which is different from the trapping of C ions for slow mode (see Figs. \ref{Fig:SlowMode}(e), \ref{Fig:SlowMode}(k)). For the slow mode, the phase velocity, $v_{\phi s}$, is nearly 3-4 times of the thermal velocity of C ions, $v_{tC}$, so C ions can be trapped by the slow mode. 
 We can find when $k\lambda_{De}=0.1, 0.3, 0.5$, the particles including electrons and H ions show a large trapping width when the driver with the linear frequency of the fast mode is off. In these cases, the linear Landau damping is relatively weak and the linear frequency is not much larger than the resonance frequency as shown in Figs. \ref{Fig:Resonance}(a), \ref{Fig:Resonance}(b) and \ref{Fig:Resonance2}(a), thus the IAW can be excited to a large amplitude especially when the driver amplitude is large. When $k\lambda_{De}=0.7$, particles including electrons, H ions and C ions can nearly not be trapped (see Figs. \ref{Fig:FastMode}(d), \ref{Fig:FastMode}(h) and \ref{Fig:FastMode}(l)). As the relative (or absolute) Landau damping rate of the fast mode, $|Im(\omega_f)/Re(\omega_f)|$ (or $|Im(\omega_f)|$), increases obviously with $k\lambda_{De}$ increasing (see Fig. \ref{Fig:LinearNonlinear}(c)) which will lead to a larger deviation between the linear frequency and the frequency calculated by the dispersion relation with no damping (shown in Fig. \ref{Fig:LinearNonlinear}(a)). In this case of $k\lambda_{De}=0.7$, the deviation between the linear frequency and the frequency calculated by the dispersion relation with no damping is nearly $7\%$ as shown in Fig. \ref{Fig:LinearNonlinear}(a) and the linear frequency is far away from the resonance frequency as shown in Fig. \ref{Fig:Resonance3}(a), thus no matter how large of the driver amplitude, the driver with the linear frequency couples very small energy into the fast mode and the IAW can nearly not be excited to a large amplitude.

To clarify the excitation of the nonlinear ion acoustic wave (IAW) by the driver with the linear frequency of the fast mode when $k\lambda_{De}$ changes, the electric field of the system for the fast mode in some cases of $k\lambda_{De}$ is given in Fig. \ref{Fig:Electric_FastMode}. In the condition of $k\lambda_{De}=0.7$ for the fast mode (shown in Fig. \ref{Fig:Electric_FastMode}(d)), when the driver is turned on, the electric field of the system will show a large response to the driver because of the internal static electric field generated by the interaction of the driving electric field and the particles, which is the same as the process in the condition of $k\lambda_{De}=0.1, 0.3, 0.5$. However, in the condition of $k\lambda_{De}=0.7$, when the driver is off, the driver with the linear frequency appears to couple small but nonzero energy into the fast mode and the amplitude of the internal static electric field will fall down to a very low level (see Fig. \ref{Fig:Electric_FastMode}(d)). For the linear Landau damping of the fast mode increases with $k\lambda_{De}$ as shown in Fig. \ref{Fig:LinearNonlinear}(c), the linear Landau damping of the fast mode in the case of $k\lambda_{De}=0.7$ is relatively stronger than that in the case of $k\lambda_{De}=0.5, 0.3$ or 0.1, which will lead to a larger deviation between the linear frequency and the frequency calculated by the dispersion relation with no damping (as shown in Fig. \ref{Fig:LinearNonlinear}(a)), as a result, the linear frequency of the fast mode is far away from the resonance frequency as shown in Fig. \ref{Fig:Resonance3}(a), which is different from that in the condition of $k\lambda_{De}=0.1, 0.3, 0.5$. As shown in Figs. \ref{Fig:Electric_FastMode}(a)- \ref{Fig:Electric_FastMode}(c), when the driver is off, the amplitude of the internal electric field in CH plasmas keeps on a relatively larger level which leads to the trapping of particles with a large trapping width. This indicates that when $k\lambda_{De}$ is not large, the nonlinear IAW can be excited by the driver with the linear frequency of the modes. However, in the condition of large $k\lambda_{De}$, especially when $k\lambda_{De}$ is as large as 0.7, the nonlinear ion acoustic wave can nearly not be excited by the driver with linear frequency of the modes.

\subsection{\label{Subsec:C}Driver with the linear frequency and the frequency calculated by the dispersion with no damping of the IAW}
\begin{figure}[!tp]
	\includegraphics[width=\columnwidth]{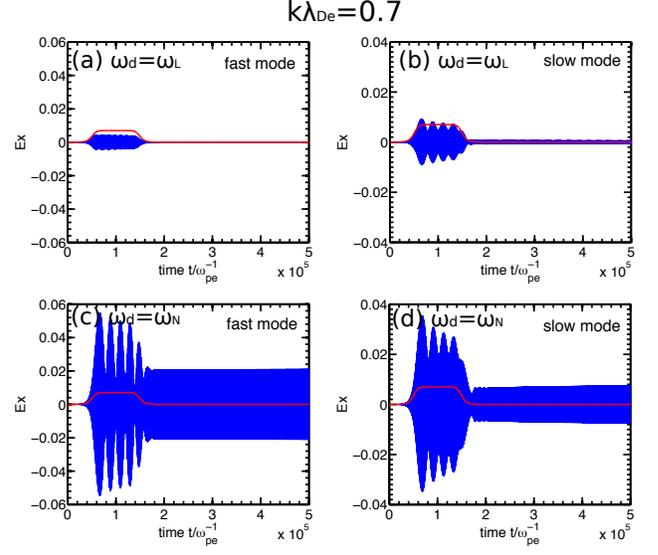}

	\caption{\label{Fig:k=0.7}(Color online) The excitation of the nonlinear ion acoustic wave by the driver with (a), (b) the linear frequency and (c), (d) the frequency calculated by the dispersion with no damping. (a), (c) are the excitation of the fast mode, $T_i/T_e=0.1$, (b), (d) are the excitation of the slow mode, $T_i/T_e=0.5$. All the cases are in the condition of $k\lambda_{De}=0.7$ and with the same driver with the maximum amplitude  $eE_d\lambda_{De}/T_e=7\times10^{-3}$. Where the vertical axis $E_x=eE\lambda_{De}/T_e$.}
\end{figure}
To verify whether the driver with the frequency calculated by the dispersion relation with no damping can excite a large amplitude nonlinear IAW especially when $k\lambda_{De}=0.7$, the cases of the driver with the frequency calculated by the dispersion with no damping and the linear frequency of the IAW are taken to excite the IAW. As shown in Fig. \ref{Fig:k=0.7}, in the condition of $k\lambda_{De}=0.7$, the driver with the linear frequency of the IAW ($\omega_d=\omega_L=0.01147\omega_{pe}$ for the fast mode, $\omega_d=\omega_L=0.01091\omega_{pe}$ for the slow mode) can nearly not excite the IAW as shown in Figs. \ref{Fig:k=0.7}(a) and \ref{Fig:k=0.7}(b). However, the driver with the frequency calculated by the dispersion with no damping of the IAW ($\omega_d=\omega_N=0.01060\omega_{pe}$ for both the fast mode and the slow mode, the frequencies of the fast mode and the slow mode in this condition are the same by coincidence, see Table \ref{table1}) can excite a large amplitude nonlinear IAW after several trapping periods as shown in Figs. \ref{Fig:k=0.7}(c) and \ref{Fig:k=0.7}(d). Although the frequency has a negative shift due to the particles trapping as shown in Fig. \ref{Fig:Resonance3}, the frequency calculated by the dispersion relation with no damping $\omega_N$ is also closer to the resonance frequency than the linear frequency $\omega_L$, thus the driver with $\omega_N$ can couple significant energy into the BGK modes as shown in Figs. \ref{Fig:k=0.7}(c) and \ref{Fig:k=0.7}(d).  This illustrates that the linear frequency of the modes is far away from the resonance frequency and the frequency calculated by the dispersion relation with no damping is closer to the resonance frequency of the nonlinear IAW than that calculated by the linear dispersion relation. If the IAW amplitude is small enough, the NFS will be very small and the resonance frequency will be very close to the frequency calculated by the dispersion relation with no damping, as shown in Fig. \ref{Fig:Resonance}.

\section{\label{Sec:Discussion}Discussion}
When the nonlinear IAW amplitude excited by the driver is weak, the nonlinear frequency shift (NFS) of the nonlinear IAW is small. The nonlinear frequency shift from particles trapping is proportional to the square root of the potential amplitude of the nonlinear IAW, $\delta \omega_N^{kin} \propto \sqrt{e\phi/T_e}$, where $\phi=E/k$, and the NFS from harmonic generation is proportional to the square of the potential amplitude of the nonlinear IAW, $\delta \omega_N^{flu} \propto (e\phi/T_e)^2$.\cite{(9)} The maximum amplitude of the driver $eE_d^{max}\lambda_{De}/T_e=1\times10^{-3}$ is taken in the cases of the resonance curve of the fast mode in the condition of $k\lambda_{De}=0.1, 0.3$ as shown in Fig. \ref{Fig:Resonance}. The corresponding peak amplitude of nonlinear IAW is very small, $eE\lambda_{De}/T_e\sim3\times10^{-4}$ in the condition of $k\lambda_{De}=0.1$ and $eE\lambda_{De}/T_e\sim6\times10^{-3}$ in the condition of $k\lambda_{De}=0.3$. In these cases, the nonlinear frequency shift is very small, which ensures that the NFS have a small effect on the resonance frequency of the nonlinear IAWs. These cases as shown in Fig. \ref{Fig:Resonance} can illustrate that the frequency calculated by the dispersion relation with no damping is much closer to the resonance frequency than the linear frequency when the NFS can be ignored. 

 However, when the large amplitude nonlinear IAW is excited by the large amplitude external driving electric field, the NFS may need to be considered. Figs. \ref{Fig:SlowMode}-\ref{Fig:Electric_FastMode} show the nonlinear IAW with large amplitude $eE\lambda_{De}/T_e\sim3\times10^{-2}$. In the condition of $k\lambda_{De}=0.1, 0.3, 0.5$, for the linear frequency $Re(\omega_L)$ is near the resonance frequency $\omega_N+\delta \omega$, the nonlinear IAW can be excited to a large amplitude. In this paper, we think the frequency calculated by the dispersion relation with no damping $\omega_N$ is the fundamental frequency, $\delta\omega$ is the quantity of the nonlinear frequency shift relative to the fundamental frequency. While in the condition of $k\lambda_{De}=0.7$, the harmonic effect is weak and the positive fluid NFS from harmonic generation is ignored. The NFS of the nonlinear IAW mainly comes from the kinetic NFS due to particles trapping, and in this condition, the kinetic NFS is negative. As shown in Fig. \ref{Fig:Resonance3}, the frequency calculated by the dispersion relation with no damping $\omega_N$ is much closer to the resonance frequency than the linear frequency $\omega_L$.
 Therefore, the driver with $\omega_N$ can excite a large amplitude IAW as shown in Figs. \ref{Fig:k=0.7}(c) and \ref{Fig:k=0.7}(d), while the driver with the linear frequecy $\omega_L$ couples very small energy into a BGK mode and can nearly not excite a large amplitude IAW no matter how large of the driver amplitude (the fast mode as shown in Figs. \ref{Fig:Electric_FastMode} (d), \ref{Fig:k=0.7} (a), the slow mode as shown in Figs. \ref{Fig:SlowMode} (b), \ref{Fig:k=0.7} (b)).

\section{\label{Sec:Summary}Summary}
The excitation of the nonlinear ion acoustic modes including the fast mode and the slow mode by the driving electric field with the linear frequency or the frequency calculated by the dispersion with no damping of the IAW when $k\lambda_{De}$ varies has been shown. For the IAW with low Landau damping can be excited preferentially, only the IAWs in the low Landau damping region are considered, such as the fast mode in the condition of $T_i/T_e=0.1$, the slow mode in the condition of $T_i/T_e=0.5$. 

When $k\lambda_{De}$ increases, the linear Landau damping, $|Im(\omega)/Re(\omega)|$, of the fast mode  increases obviously in the region of low Landau damping ($T_i/T_e\lesssim0.2$). This provides the possibility of the suppression of SBS in the region of large wave number , i.e., the region of high temperature and low density of the electrons. However, this is the linear result. Considering the particle trapping, the distribution of particles will keep flat at the phase velocity, thus turning off Landau damping. As a result, the dispersion relation with no damping is provided to calculate the frequency of the nonlinear IAWs. When $k\lambda_{De}$ increases, the frequency calculated by the linear dispersion relation and that calculated by the dispersion relation with no damping deviate each other larger and larger. Especially, in the condition of $k\lambda_{De}=0.7$ for the fast mode, the deviation is as large as nearly $7\%$. 

By Vlasov simulation, the resonance curves show that the resonance frequency is much closer to the frequency calculated by the dispersion relation with no damping than that calculated by the linear dispersion relation. When $k\lambda_{De}$ is not large, such as $k\lambda_{De}=0.1, 0.3, 0.5$, the frequency of the linear fast mode in the condition of $T_i/T_e=0.1$ and the linear slow mode in the condition of $T_i/T_e=0.5$ are near the resonance frequency of the nonlinear IAW modes, thus, the nonlinear IAW can be excited by the driver with the linear frequency of the modes. However, in the condition of $k\lambda_{De}=0.7$, for the linear frequency is far away from the resonance frequency of the nonlinear IAW, the nonlinear IAW can nearly not be excited by the external driver with the linear frequency of the IAW (no matter how large of the amplitude of the driver). While the driver with the frequency calculated by the dispersion relation with no damping can excite the large amplitude nonlinear IAW.

Our findings indicate: The frequency calculated by the dispersion relation with no damping is much closer to the resonance frequency of the small amplitude nonlinear IAW than the linear frequency. When $k\lambda_{De}$ is large, such as $k\lambda_{De}=0.7$, the linear frequency can not be applied to exciting the nonlinear IAW, while the frequency calculated by the dispersion relation with no damping can be applied to exciting the nonlinear IAW.

\begin{acknowledgments}
We are pleased to acknowledge useful discussions with L. H. Cao. This research was supported by the National Natural Science Foundation of China (Grant Nos. 11575035, 11475030 and 11435011) and National Basic Research
Program of China (Grant No. 2013CB834101).
\end{acknowledgments}



\end{document}